\documentstyle[12pt,graphics,pra,aps]{revtex}
\input{epsf}
\begin{document}
\title{Absorption with inversion and amplification without 
inversion in a coherently prepared V - system: a dressed state approach}
	
\author{ D. Braunstein and R. Shuker* }

\date{\today}
\address{ Department of Physics, Ben Gurion University of the Negev, 84105
Beer Sheva, Israel\\
}
\maketitle
\begin{abstract}
Light induced absorption with population inversion and
 amplification without population inversion (LWI) in a 
 coherently prepared closed three level V - type system are 
 investigated.
This study is performed from the point of view of a two color 
dressed state basis. Both of these processes are possible due to 
 atomic coherence and quantum interference  contrary to simple intuitive 
predictions.
Merely on physical basis, one would expect a complementary process to 
the amplification without inversion. We believe that absorption in the 
presence of population inversion found in the dressed state picture 
utilized in this study, constitutes such a process.

 Novel approximate analytic time dependent solutions, 
for coherences and populations are obtained, and are compared 
with full numerical solutions exhibiting excellent agreement.
Steady state quantities are also 
calculated, and the conditions under which the system exhibits 
absorption and gain 
 with and without inversion, in the dressed state representation are derived. 
It is found that for a weak probe laser field absorption with 
inversion and
amplification  without inversion may occur, for a range of system 
parameters.
\end{abstract}

\thanks *shuker@bgumail.bgu.ac.il
\newpage
\section{Introduction}

Recently there has been tremendous interest in the study of light 
amplification and lasing without the requirement of population 
inversion (LWI), potentially capable of extending the range of laser 
devices to spectral region in which, for various reasons, population 
inversion is difficult to achieve. These spectral regions include UV 
which can be 
obtained from atomic vapor, and mid-to-far infrared, obtained by 
interssubband transitions in quantum wells. Many models for LWI have 
been proposed, mostly three and four level schemes, in $\Lambda$ and 
$V$ configurations. The dependence of  optical gain on 
system parameters has also been investigated \cite{1,2,3,4,5,6,7,8,9,%
10,11,12,13,14,15,16,17}. \\
The key mechanism, which is common to most of the proposed schemes, is 
the utilization of external coherent fields, that induce quantum 
coherence and interference in multilevel systems. An exception 
of LWI without the use of coherent fields was also reported \cite{9}. 
In particular it was shown that if atomic coherence between 
certain atomic states is established , different absorption processes may interfere 
destructively leading to the reduction or even the cancellation of 
absorption \cite{7,13}. At the same time stimulated emission may remain intact, 
leading to the possibility of gain even if only a small fraction of 
the population is in the excited state. \\
Experimental observations of inversionless gain and lasing without 
inversion have been reported by several groups \cite{18,19,20,21}. Inversionless 
lasers have been shown to have unique properties such as non 
classical photon statistics and substantially narrow spectral 
features \cite{22,23}. In a recent paper, Y. Zhu \cite{12} analyzed the 
transient and steady state properties of light amplification without 
population inversion in a closed three level V type system in the 
bare state basis. Steady state dressed state populations were also 
calculated, in the limit of strong driving laser. \\

In this paper we study a V- type three level model within the framework of the 
dressed state basis, and give explicit analytic time dependent solutions, as 
well as steady state solutions for populations and coherences. 
This paper details the  calculations  of LWI in the 
dressed state basis, which is valid to some extent for atoms dressed 
both by the pump and probe lasers. More over we offer a set of 
theoretical tools allowing one to obtain explicit time dependent 
solutions for populations and coherences.
These calculations show the possibility of inversion and 
inversionless gain in the dressed basis as well as gain without 
population inversion in the bare state representation.
A novel interesting feature found in this study 
is the existence of absorption 
despite population inversion. Although this effect is contrary to 
simple physical intuitive explanation of absorption, this process has 
a conceptual reasoning. From basic physical arguments one expects a 
complementary process to amplification without inversion. We believe 
that absorption in the presence of population inversion, found in the 
dressed state picture, constitutes such a 
process. It is another manifestation of the quantum interference that 
may occur in multilevel systems where coherently prepared states 
present the possibility of interfering channels \cite{23a}. This phenomenon should be 
interpreted as a  quantum  interference  constructive  process
for the stimulated absorption just as LWI is obtained from a quantum 
 interference {\it{destructive}} process of absorption.

 In Sec. {\bf{II}} we present the 
model system and the master equation used to derive the equations of 
motion for the elements of the density matrix. We have chosen to 
employ a fully quantum mechanical Hamiltonian, even though later on 
the density matrix equations are reduced to their semiclassical 
version. The quantum mechanical Hamiltonian gives rise to a simple 
picture of the stationary dressed states. In Sec. {\bf{III}} the 
dressed levels are introduced and the master equation is projected 
over the dressed state basis. Physical interpretation of 
relaxation coefficients in the dressed basis is given and the role that quantum coherences and
interferences play is elucidated. In Sec. {\bf{IV}} we present time 
dependent and steady state solutions for the dressed state populations 
and coherences. Comparison is made with the bare state results. 
\newpage 
\section{Hamiltonian and master equation for the atom}
\begin{center}
\begin{figure}[p]
	\centering
	\includegraphics{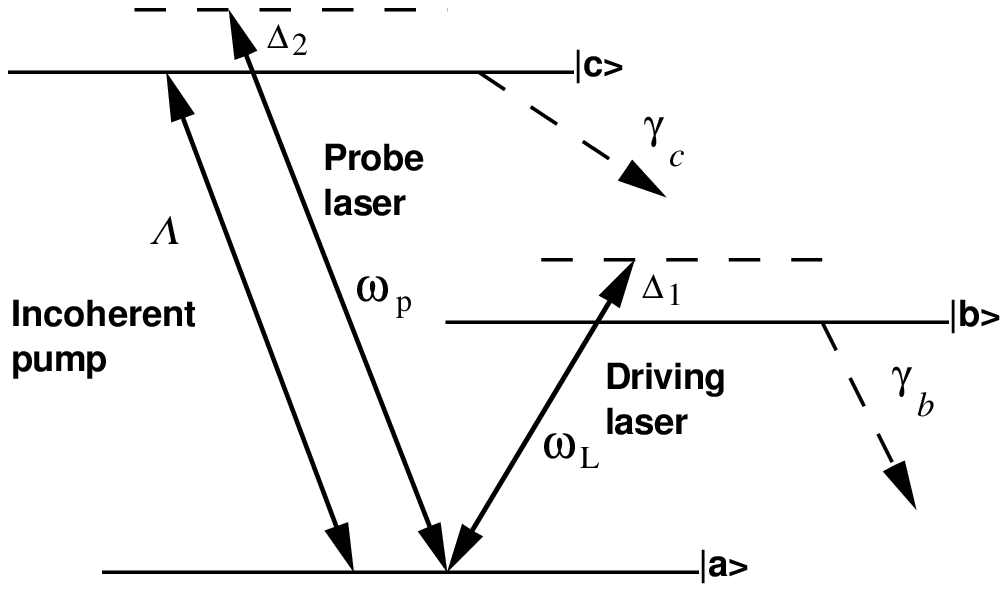}
\caption{A three level V - type system for LWI}
	\label{f1}
\end{figure}
\end{center}

Let us consider the closed V-type three level system illustrated in 
figure \ref{f1}. The transition $|a>\leftrightarrow|b>$ of frequency 
$\omega_{ba}$ is driven by a strong 
single mode laser of frequency $\omega_{L}$€. The transition $|a>\leftrightarrow|c>$ 
of frequency $\omega_{ca}$ € is pumped incoherently with a rate $ \Lambda $. 
A single mode probe laser(not necessarily weak) is applied to the 
transition $|a>\leftrightarrow|c>$. $\gamma_{b}$€ ($\gamma_{c}$€) is 
the spontaneous emission rate from the state $|b>$ ($|c>$). The 
states $|b>$ and $|c>$ are not directly coupled.\\

 We have chosen to work within the frame of the master 
equation for the atom, since it being an operator equation independent 
of representation, it can be projected over any basis. We use a 
generalization of a standard master equation adjusted to account for 
the scheme described above \cite{24}. The master equation is given by:

\begin{eqnarray}
	\dot{\sigma} & = & -\frac{i}{\hbar}[ 
	H,\sigma]-\frac{\gamma_{b}€}{2}€[s_{+}€s_{-}€\sigma+\sigma 
s_{+}€s_{-}€]+\gamma_{b}€s_{-}€ \sigma s_{+}€
 -\frac{\gamma_{c}€}{2}€[s'_{+}€s'_{-}€\sigma+\sigma
 s'_{+}€s'_{-}€] 
 +\gamma_{c}€s'_{-}€\sigma s'_{+}€\nonumber 
\\
	& & \mbox{}-	\frac{\Lambda}{2}€[s'_{+}€s'_{-}€\sigma+\sigma 
	s'_{+}€s'_{-}€]-\frac{\Lambda}{2}€[s'_{-}€s'_{+}€\sigma+\sigma 
	s'_{-}€s'_{+}€]+\Lambda s'_{+}€\sigma s'_{-}€+\Lambda s'_{-}€\sigma s'_{+},
	\label{1}
\end{eqnarray}€

here $\sigma$ is the density operator for the atom. $s_{+}€(s_{-})€$, $s'_{+}€(s'_{-})€$ are the atomic raising and 
lowering operators, for the $|a>\leftrightarrow|b>$ and 
$|a>\leftrightarrow|c>$ transitions respectively. {\it{H}} is the 
Hamiltonian of the global system and we take it to be fully quantum 
mechanical. The quantum mechanical Hamiltonian gives rise to a simple 
picture of the stationary dressed states. The Hamiltonian in the 
dipole and rotating wave approximation is given by:
\begin{eqnarray}
H & = &\hbar\omega_{ba}€|b><b|+\hbar\omega_{ca}€|c><c|+\hbar\omega_{L}€(a^{\dagger}€a 
+\frac{1}{2}€)+\hbar\omega_{p}€(a'^{\dagger}€a'+\frac{1}{2}€) \nonumber	  \\
& & \mbox{}+g (s_{+}€a+s_{-}a^{\dagger})+g'(s'_{+}a'+s'_{-}a'^{\dagger}).
\label{2}
\end{eqnarray}€
$g$ and $g'$ are coupling constants and are assumed to be real. The 
eigenstates of 
the unperturbed part of the Hamiltonian form a three dimensional 
manifold labeled~ by the atomic indexes, the laser photon number $N$ and by 
the probe photon number $N'$. The manifold is written
\begin{equation}
\varepsilon (N,N')=\{|a,N+1,N'+1>,|b,N,N'+1>,|c,N+1,N'>\}.
	\label{3}
\end{equation}€
 
 We represent the uncoupled  eigenstates of the atom and the two 
 non-interacting field modes as:
\begin{equation}
	|a,N+1,N'+1>=\left (\begin{array}{c}
	1\\0\\0
	\end{array} \right )\hspace{.15in}€
	|b,N,N'+1>=\left (\begin{array}{c}
	0\\1\\0
	\end{array} \right )\hspace{.15in }€
	|c,N+1,N'>=\left (\begin{array}{c}
	0\\0\\1
	\end{array} \right ).
	\label{4}
\end{equation}€

In this basis the  hamiltonian takes the form
\begin{equation}
	H_{Int}=\left ( \begin{array}{ccc}
	0 &-\Omega & -G \\
	-\Omega & -\Delta_{1} & 0 \\
	-G & 0 & -\Delta_{2}
	\end{array} \right ).
	\label{5}
\end{equation}

Where we have defined the Rabi frequencies and the detunigs, in their 
quantum form by 

\begin{equation}
	-g \sqrt{N+1}=\hbar \Omega \hspace{.1in } ; \hspace{.1in }-g' \sqrt{N'+1}=\hbar 
G \hspace{.1in };\hspace{.1in } \Delta_{1}=\omega_{L}-\omega_{ba}\hspace{.1in }
 ;\hspace{.1in } 
\Delta_{2}=\omega_{p}-\omega_{ca}.
\end{equation}

To obtain the semiclassical equations of motion for the elements of 
the density matrix we project the master equation (\ref{1}) over the 
basis (\ref{4}) and perform the following reduction operation by 
introducing the reduced populations and coherences $\rho_{ij}$ via:
 
 \begin{mathletters}
\begin{eqnarray}
\rho_{aa} & = & \sum_{N,N'}<a,N+1,N'+1|\sigma |a,N+1,N'+1>,
	\label{6a}  \\
\rho_{ab} & = &  \sum_{N,N'}<a,N+1,N'+1|\sigma |b,N,N'+1>. 
\label{6b}
\end{eqnarray}
\end{mathletters}
and similar relations for the other populations and coherences. Taking 
into account the above reduced quantities we obtain the  
density matrix equations

\begin{mathletters}
\begin{eqnarray}
	\dot\rho_{aa} & = & - \Lambda \rho_{aa}+(\Lambda +\gamma_{c}) 
	\rho_{cc}+\gamma_{b} \rho_{bb}+i\Omega (\rho_{ba}-\rho_{ab})+i G 
	(\rho_{ca}-\rho_{ac}), 
	\label{7a} \\
\dot\rho_{bb} & = & -\gamma_{b}\rho_{bb}+i\Omega 
(\rho_{ab}-\rho_{ba}) ,
\label{7b}\\
	\dot\rho_{cc} & = & \Lambda \rho_{aa}-(\Lambda 
	+\gamma_{c})\rho_{cc}+i G(\rho_{ac}-\rho_{ca}),
	\label{7c} \\
	\dot\rho_{ab} & = & -[\frac{1}{2} (\Lambda 
	+\gamma_{b})+i\Delta_{1}]\rho_{ab}+i \Omega (\rho_{bb}-\rho_{aa})+i 
	G \rho_{cb} ,
	\label{7d}\\
	\dot\rho_{ac} & = & -[\frac{1}{2}( \Lambda+\gamma_{c})+i 
	\Delta_{2}]\rho_{ac}+i G(\rho_{cc}-\rho_{aa})+i\Omega \rho_{bc},
 \label{7e} \\
	\dot\rho_{bc} & = & 	
	-[\frac{1}{2}(\Lambda+\gamma_{b}+\gamma_{c})+i(\Delta_{2}-\Delta_{1})]\rho_{bc}+
	i \Omega \rho_{ac}-i G \rho_{ba}.
	\label{7f}
\end{eqnarray}
\end{mathletters}
\newpage
\section{Dressed states and density matrix equations in the dressed 
state basis}€
\paragraph*{} The dressed states are obtained by finding the 
the eigenvectors of the interaction Hamiltonian, Eq.~(\ref{5}). To simplify things a 
little we take both the driving laser field and the probe field to be in exact 
resonance with the corresponding bare state transitions, i.e., we 
take $\Delta_{1}=~\Delta_{2}=0$. When the detunings are set to  zero 
we notice that the energies in the bare state basis are all 
degenerate, and in fact equal to zero (in the interaction picture). 
Carrying out the diagonalization procedure we obtain the following 
eigenstates:
\begin{mathletters}
\begin{eqnarray}
	|\alpha (N) (N')> & = & -\frac{G}{R}\hspace{.1 cm} |b,N,N'+1>+\frac{\Omega}{R}\hspace{.1 cm}
	|c,N+1,N'>,
	\label{8a}\\
	|\beta (N) (N')> & = & -\frac{1}{\sqrt{2}}\hspace{.1 cm}
	|a,N+1,N'+1>+\frac{\Omega}{\sqrt{2}R}\hspace{.1 cm}
	|b,N,N'+1>+\frac{G}{\sqrt{2}R}\hspace{.1 cm}|c,N+1,N' >,
	\label{8b} \\
	|\gamma (N) (N')> & = & \frac{1}{\sqrt{2}}\hspace{.1 cm}|a,N+1,N'+1>+\frac{\Omega}{\sqrt{2}R} 
	\hspace{.1 cm}|b,N,N'+1>+\frac{G}{\sqrt{2}R}\hspace{.1 cm} |c,N+1,N'>.
	\label{8c}
\end{eqnarray}
 \end{mathletters}
 
and the corresponding energies 

\begin{equation}
	E_{|\gamma>}=-\hbar R\hspace{.5 in};\hspace{.5 in}E_{|\alpha>}=0\hspace{.5 
	in};\hspace{.5 in}E_{|\beta>}=\hbar R.
	\label{9}
\end{equation}

Where we have introduced the on resonance, two field generalized Rabi flopping frequency 
$R=\sqrt{\Omega^{2}+G^{2}}$.

 \paragraph*{}Note that for the case $G=0$ equations (\ref{8b}) and 
 (\ref{8c}) render 
the usual coupling and non coupling dressed states, while the 
$|\alpha>$ state which is identical to the $|c>$ state is not involved in 
the interaction altogether.
 The energy ladder is shown schematically in figure 2. We can see that 
 one state remained intact, while the other two states were displaced
  by an energy amount of $\hbar R$, with respect to the bare 
 states. In the strong driving field limit, i.e., when $\Omega>>~G$, we 
 see that the $|\alpha (N) (N')>$ state has the character of the 
 excited bare state $|c,N+1,N'>$ and hence is expected to be less 
 populated than the other two states. By contrast, the $|\beta (N) 
 (N')>$ and $|\gamma (N) (N')>$ states have a ground state character, 
 and hence will be more populated than the $|\alpha (N) (N')>$ state. 
 However both $|\beta>$ and $|\gamma>$ states are also contaminated by the same amount of the first 
 excited level and thus they are expected to possess the same population 
 content. 
 \paragraph*{} The eigenstates of equations (\ref {8a})-(\ref{8c}) define a rotation 
 matrix (transformation matrix)
 \begin{equation}
 	T=\left ( \begin{array}{ccc}
	0 &-\frac{G}{R} & \frac{\Omega}{R} \\
	\\ 
	-\frac{1}{\sqrt{2}} & \frac{\Omega}{\sqrt{2}R} & \frac{G}{\sqrt{2}R} \\
	\\ 
	\frac{1}{\sqrt{2}} & \frac{\Omega}{\sqrt{2}R} & \frac{G}{\sqrt{2}R}
	\end{array} \right )
 	\label{10}
 \end{equation}€
 
  that diagonalizes the Hamiltonian of 
 Eq.(\ref {5}) via the matrix product $T H T^{-1}$. 
 Thus the density 
 operator in the dressed atom basis, $\rho^{Dr}$, will be given by the matrix product
 \begin{equation}
     	\rho^{Dr}=T \rho^{B} T^{-1}
     	\label{11}
     	\end{equation}
where $\rho^{B}$ is the density operator in the bare basis.

\paragraph*{}Projection of the master equation over the dressed state 
basis yields particularly simple equations for the first part of 
Eq.(\ref {1}), i.e., the Hamiltonian part of the master equation. 
However, in the dressed atom  basis the relaxation terms of Eq.(\ref{1})
 give rise to equations that are not as simple as equations 
(\ref{7a}-\ref{7f}). 
In particular couplings between dressed state populations and 
coherences between two dressed states appear. In the next section we 
present an approximate version of the complete set of equations 
given bellow.

 The equation of motion for the density matrix elements 
in the dressed state representation are given by:
\begin{mathletters}
 \begin{eqnarray}
 	 \dot\rho_{\alpha\alpha }& = & -(\Gamma_{\alpha}+\Lambda') 
 	 \rho_{\alpha \alpha}+\tilde{\Gamma}(\rho_{\alpha \beta}+\rho_{\beta \alpha}) 
 	 +\tilde{\Gamma}(\rho_{\alpha \gamma}+\rho_{\gamma \alpha})+ 
 \frac{1}{2}\Lambda' \rho_{\beta 
 	 \beta}\nonumber \\
 	 & &\mbox{}-\frac{1}{2}\Lambda'(\rho_{\beta \gamma}+\rho_{\gamma 
 	 \beta})+\frac{1}{2}\Lambda' \rho_{\gamma \gamma} 
 	 \label{12a}\\
 \dot \rho_{\alpha \beta} & = & -(\Gamma_{\alpha 
 \beta}-iR)\rho_{\alpha 
 \beta}-(\Gamma_{\beta}-\frac{1}{4}\Lambda')\rho_{\alpha \gamma 
 }-\tilde{\Gamma'}\rho_{\beta 
 \gamma}+(\tilde{\Gamma}-\tilde{\Gamma'})\rho_{\gamma \beta}\nonumber  \\
 & & \mbox{}+ \tilde{\Gamma}\rho_{\alpha 
 \alpha}+(\tilde{\Gamma}+\tilde{\Gamma'})\rho_{\beta 
 \beta}+\tilde{\Gamma'}\rho_{\gamma \gamma}, 
 \label{12b}\\
 	\dot\rho_{\alpha \gamma } & = & -(\Gamma_{\alpha 
 	\beta}+iR)\rho_{\alpha 
 	\gamma}-(\Gamma_{\beta}-\frac{1}{4}\Lambda')\rho_{\alpha 
 	\beta}+(\tilde{\Gamma}-\tilde{\Gamma'})\rho_{\beta 
 	\gamma}-\tilde{\Gamma'}\rho_{\gamma \beta}\nonumber \\
 	& & \mbox{}+ \tilde{\Gamma}\rho_{\alpha \alpha}+\tilde{\Gamma'}\rho_{\beta 
 	\beta}+(\tilde{\Gamma}+\tilde{\Gamma'}) \rho_{\gamma \gamma}
 	\label{12c}  \\
 	\dot \rho_{\beta \beta } & = & 
 	-(\Gamma_{\beta}+\frac{1}{2}\Lambda )\rho_{\beta 
 	\beta}+\frac{1}{2}(\Gamma_{\alpha}+\Lambda') \rho_{\alpha 
 	\alpha}+(\Gamma_{\beta}+\frac{1}{2}\Lambda 
 	-\frac{1}{2}\Lambda')\rho_{\gamma \gamma}\nonumber \\
 	& & \mbox{}- \tilde{\Gamma}(\rho_{\alpha \gamma}+\rho_{\gamma 
 	\alpha})+\frac{1}{4}\Lambda'(\rho_{\beta \gamma}+\rho_{\gamma 
 	\beta}),\label{12d} \\
 	\dot \rho_{\beta \gamma} & = & -(\Gamma_{\beta 
 	\gamma}+2iR)\rho_{\beta \gamma}-(\Gamma_{\beta}+\frac{1}{2}\Lambda 
 	-\frac{1}{2}\Lambda' )\rho_{\gamma 
 	\beta}+\tilde{\Gamma}(\rho_{\alpha \beta}+\rho_{\gamma 
 	\alpha})\nonumber\\
 	&& \mbox{}+ 2\tilde{\Gamma}(\rho_{\beta \alpha}+\rho_{\alpha \gamma}) 
 	-\frac{1}{2}(\Gamma_{\alpha}+\Lambda')\rho_{\alpha 
 	\alpha}-(2 \Gamma_{\beta}-\frac{1}{4}\Lambda')(\rho_{\beta 
 	\beta}+\rho_{\gamma \gamma })\label{12e}\\
 	\dot \rho_{\gamma \gamma} & = & 
 	-(\Gamma_{\beta}+\frac{1}{2}\Lambda)\rho_{\gamma 
 	\gamma}-\tilde{\Gamma}(\rho_{\alpha \beta}+\rho_{\beta 
 	\alpha})+\frac{1}{4}\Lambda'(\rho_{\beta \gamma}+\rho_{\gamma 
 	\beta}) \nonumber \\
 	&& \mbox{}+ \frac{1}{2}(\Gamma_{\alpha}+\Lambda')\rho_{\alpha \alpha}+ 
 	(\Gamma_{\beta}+\frac{1}{2}\Lambda -\frac{1}{2}\Lambda')\rho_{\beta 
 	\beta}\label{12f}
 \end{eqnarray}€ 
\end{mathletters}
Where we have again made use of the reduction operation 

\begin{equation}
	\rho_{i,j}=\sum_{N,N'}^{€}<i (N) (N')|\sigma | j (N) (N')>\hspace{1 cm} 
i,j=|\alpha>, |\beta>, |\gamma>
	\label{13}
\end{equation}€

In obtaining equations (\ref{12a}) - (\ref{12f}) we have introduced 
the following notation

\begin{eqnarray*}
	 \Gamma_{\alpha}&=& \frac{1}{R^{2}}(\gamma_{b}G^{2}+\gamma_{c}\Omega^{2})
	  \hspace{.2cm}, \hspace{.2cm}\Gamma_{\beta}=\frac{1}{4R^{2}}
	  (\gamma_{b}\Omega^{2}+\gamma_{c}G^{2})\hspace{.2cm},\\ 
	   \Gamma_{\alpha \beta}&=&\Gamma_{\beta}+
	   \frac{1}{2}(\Gamma_{\alpha}+\Lambda+\Lambda'/2)\hspace{.2cm},\hspace{.2cm}
\Gamma_{\beta\gamma}=3\Gamma_{\beta}+\frac{3}{2}\Lambda 
-\Lambda',\\ 
\tilde{\Gamma}&=&\frac{G\Omega}{2\sqrt{2}R^{2}}
(\gamma_{b}-\gamma_{c}-\Lambda)\hspace{.2cm},\hspace{.2cm}
 \tilde{\Gamma'}=\frac{G\Omega\Lambda}{2\sqrt{2}R^{2}} \hspace{.2cm}, \hspace{.2cm}
 	\Lambda' =\frac{\Lambda\Omega^{2}}{R^{2}}.
\end{eqnarray*}

$\Gamma_{\alpha}$, $\Gamma_{\beta}$ are the spontaneous emission decay 
rate of the $|\alpha>$, $|\beta>$ and $|\gamma>$ states ($|\gamma> $ 
also
decays with a rate $\Gamma_{\beta}$). More precisely the 
state $|\alpha(N)(N')>$ decays by spontaneous emission with a rate 
$\Gamma_{\alpha}$ to the levels $|\beta(N-1)(N')>$, 
$|\gamma(N-1)(N')>$ , $|\beta(N)(N'-1)>$ and 
$|\gamma(N)(N'-1)>$. Similarly the levels $|\beta(N)(N')>$, 
$|\gamma(N)(N')>$ decay with the same rate $\Gamma_{\beta}$ to the 
same levels as $|\alpha(N)(N')>$. The coherences $\rho_{\alpha 
\beta}$, and
 $\rho_{\alpha \gamma}$ ($\rho_{\beta \gamma}$) decay with a rate 
 $\Gamma_{\alpha \beta}$ ($\Gamma_{\beta \gamma}$). $\Lambda'$ is 
a dressed picture pump rate which causes population and depopulation of the 
 dressed levels. It also has an important influence on
 the coherences as can readily be 
 seen from Eqs. (\ref{12a}) - (\ref{12f}). $\tilde{\Gamma}$ and $\tilde{\Gamma'}$ 
 are identified as  interference terms. They involve the product of two Rabi 
 frequencies. Both parameters vanish whenever either $G$ or $\Omega$ are 
 zero. These terms are responsible for the amplification without 
 inversion and for the absorption despite the inversion. This fact is
 verified numerically. When we have set 
 both $\tilde{\Gamma}$ and $\tilde{\Gamma'}$ zero (this happens when $\Lambda=0$ and $
 \gamma_{b}=\gamma_{c}$) any previously obtained gain has vanished. 
 The first and second terms in $\Gamma_{\alpha \beta}$ describe 
 damping of the atomic coherence due to radiative transitions of the 
 levels involved to lower levels, and is equal to half  the sum of 
 all transition rates starting from $|\alpha(N)(N')>$ and 
 $|\beta(N)(N')>$. The third term in $\Gamma_{\alpha \beta}$ describes 
 coherence damping due to the incoherent pump. The interpretation of 
 $\Gamma_{\beta \gamma}$ is similar except that the $3\Gamma_{\beta}$ 
 is composed of a $2\Gamma_{\beta}$ term responsible for the coherence damping 
 due to radiative transition, plus a single $\Gamma_{\beta}$ component 
 resulting from transfer of coherence from higher levels belonging to 
 higher manifolds. This fact would have been transparent had  we 
 written the non reduced version of Eqs. ( \ref{12a}) - (\ref{12f}).
 Inspection of equations (13) reveals that $\rho_{\alpha \beta}$ and 
  $\rho_{\alpha \gamma}$ have the same free frequency R, however they 
  oscillate out of phase. The free evolution frequency of $\rho_{\beta 
  \gamma}$ is twice as large, as both levels $|\beta>$ and $|\gamma>$
 are contaminated by the bare ground state $|a>$. Note that the closure of the system is 
  satisfied by Eq.(\ref{12a})-(\ref{12f}), i.e.,
  $\frac{d}{dt}(\rho_{\alpha \alpha}+ \rho_{\beta 
  \beta}+\rho_{\gamma\gamma})=0$.
  Gain or absorption coefficient for the $|j> \rightarrow |i>$ 
  transition is proportional to $Im[\rho_{ij}]$. In our notation 
  amplification will occur if $Im[\rho_{ij}]>0 $. \\
  In the next section we present approximate solutions
  of Eq.s (\ref{12a})-(\ref{12f}), both the temporal 
  and the steady state cases. These will be compared with numerical 
  calculations of the full system, i.e., without any approximation. 

\section{Density matrix equations in the dressed state basis in the 
secular approximation}
As mentioned before the Hamiltonian part of the master equation has a 
simple form in the dressed  state basis given by Eqs. 
(\ref{8a}-\ref{8c}) (the Hamiltonian is diagonalized in the dressed 
state representation). The problem arises when the spontaneous 
emission and pump terms are present in the master equation,Eq.(\ref{1}), giving the complicated 
couplings appearing in equations (\ref{12a}) - (\ref{12f}). Solving 
exactly 
the complete set seems to be a formidable task even with ``Mathematica'' \cite{25}.
However, the situation 
can be simplified if the frequency difference
between the dressed 
states of the manifold, namely the Rabi flopping frequency $R$
 is large compared with the rates $\gamma_{b}$, $ \gamma_{c}$ 
,$\Lambda$. We can then ignore the ``nonsecular'' terms, i.e., 
couplings between populations and coherences (see \cite{24}).

\paragraph{The evolution of the population terms.} When the ``nonsecular'' couplings 
between populations and coherences 
are ignored we obtain from (\ref{12a}), (\ref{12d}) and (\ref{12f}) the 
following equations for the populations:

\begin{mathletters}
\begin{eqnarray}
\dot \rho_{\alpha \alpha} & = & -(\Gamma_{\alpha}+\Lambda') 
\rho_{\alpha \alpha}+\frac{1}{2}\Lambda'\rho_{\beta\beta} 
+\frac{1}{2}\Lambda'\rho_{\gamma \gamma},
	\label{13a}  \\
	\dot\rho_{\beta\beta} & = & -(\Gamma_{\beta}+\frac{1}{2}\Lambda)\rho_{\beta\beta}+
	\frac{1}{2}(\Gamma_{\alpha}+\Lambda')\rho_{\alpha \alpha}+	
	(\Gamma_{\beta}+\frac{1}{2}\Lambda-\frac{1}{2}\Lambda')\rho_{\gamma\gamma},
	\label{13b}  \\
\dot\rho_{\gamma\gamma} & = & 
-(\Gamma_{\beta}+\frac{1}{2}\Lambda)\rho_{\gamma\gamma}+
	\frac{1}{2}(\Gamma_{\alpha}+\Lambda')\rho_{\alpha \alpha}+	
	(\Gamma_{\beta}+\frac{1}{2}\Lambda-\frac{1}{2}\Lambda')\rho_{\beta\beta}.
	\label{13c}
\end{eqnarray}
\end{mathletters}

Note that population conservation is still maintained.
The interpretation of equations (\ref{13a})-(\ref{13c}) is very 
clear. The state $|\alpha>$ is depopulated with a rate 
$(\Gamma_{\alpha}+\Lambda')$, which in turn, is distributed equally among the 
states $|\beta>$ and $|\gamma>$, as can be readily seen from the one 
half factor multiplying the coefficient of  $\rho_{\alpha \alpha}$ in 
(\ref{13b}) and (\ref{13c}). The state $|\alpha>$ is also being 
populated with a rate $\frac{1}{2}\Lambda$ by the states $|\beta>$ 
 and $|\gamma>$. The state $|\beta>$ ($|\gamma>$) is depopulated at a rate 
$(\Gamma_{\beta}+\frac{1}{2}\Lambda)$ and repopulated with the same 
rate from $|\gamma>$ ($|\beta>$).
The set (\ref{13a}) -(\ref{13c}) can be solved exactly by calculating its 
eigenvalues and eigenstates, subject to the condition 
$\rho_{\alpha\alpha}+\rho_{\beta\beta}+\rho_{\gamma\gamma}=1$. This 
yields the temporal solution

\begin{mathletters}
\begin{eqnarray}
	\rho_{\alpha\alpha}(t) & = & 
	\rho_{\alpha\alpha}^{st}+[\rho_{\alpha\alpha}(0)-\rho_{\alpha\alpha}^{st}]
	e^{-(\Gamma_{\alpha}+\frac{3}{2}\Lambda')t},
	\label{14a}  \\
\rho_{\beta\beta}(t) & = & \rho_{\beta\beta}^{st}+[\rho_{\beta\beta}(0)-
\rho_{\beta\beta}^{st}+\frac{1}{2}(\rho_{\alpha\alpha}(0)-\rho_{\alpha\alpha}^{st})]
e^{-(2\Gamma_{\beta}+\Lambda-\frac{1}{2}\Lambda')t}\nonumber \\
&-&\frac{1}{2}[\rho_{\alpha\alpha}(0)
-\rho_{\alpha\alpha}^{st}]e^{-(\Gamma_{\alpha}+\frac{3}{2}\Lambda')t},
	\label{14b}  \\
\rho_{\gamma\gamma}(t) & = & \rho_{\gamma\gamma}^{st}-[\rho_{\beta\beta}(0)-
\rho_{\beta\beta}^{st}+\frac{1}{2}(\rho_{\alpha\alpha}(0)-\rho_{\alpha\alpha}^{st})]
e^{-(2{\Gamma_{\beta}}+\Lambda-\frac{1}{2}\Lambda')t} \nonumber \\
&-&\frac{1}{2}[\rho_{\alpha\alpha}(0)
-\rho_{\alpha\alpha}^{st}]e^{-(\Gamma_{\alpha}+\frac{3}{2}\Lambda')t}.
	\label{14c}
\end{eqnarray}
\end{mathletters}

where $\rho_{ii}(0)$ are the initial populations. The steady state 
populations are given by

\begin{mathletters}
\begin{eqnarray}
	\rho_{\alpha\alpha}^{st} & = 
	&\frac{\Lambda'}{2\Gamma_{\alpha}+3\Lambda'},
	\label{15a}  \\
	\rho_{\beta\beta}^{st}=\rho_{\gamma\gamma}^{st} & = & 
	\frac{\Gamma_{\alpha}+\Lambda'}{2\Gamma_{\alpha}+3\Lambda'}.
	\label{15b}
\end{eqnarray}
\end{mathletters}

We can see that the population $\rho_{\beta\beta}$, $\rho_{\gamma\gamma}$ 
have similar 
behavior, though not identical, a fact which is not surprising at 
all in light of the very similar composition of the states $|\beta>$ 
and $|\gamma>$. The population of the state $|\alpha>$ is unique in the sense that 
it decays with only one decay constant, while the other populations 
have a composite decay.
 Figure 3 shows a comparison between the exact 
 solution for $\rho_{\alpha\alpha}$ and $\rho_{\beta \beta}$ 
(solid line), obtained by solving numerically the equation set 
(\ref{12a}) -(\ref{12f}), with our approximate analytic solution  (\ref{14a}) - 
(\ref{14c}) shown in dashed line.
The normalized parameters for the numerical simulation were set to be
$\Omega=20\gamma_{c}$, $\gamma_{b}=2\gamma_{c}$, $G=0.1\gamma_{c}$, 
 $\Lambda=3\gamma_{c}$.
We can see that the population  
$\rho_{\alpha\alpha}$ is a monotonically  increasing, oscillating 
function of time which reaches a steady state value 
$\rho_{\alpha\alpha}^{st}\approx 0.27 $. The behavior of $\rho_{\beta\beta}$ is 
 opposite, i.e., it is a monotonically decreasing oscillating function 
and it reaches the steady state value $\rho_{\beta\beta}^{st}\approx 0.36 $.
$\rho_{\gamma\gamma}$ is not shown because its similarity to 
$\rho_{\beta\beta}$, due the choice of parameters made.
The approximate solutions describe nicely the envelope of oscillation 
and the correct expression for the steady state. One can see that 
$\rho_{\alpha\alpha}<\rho_{\beta\beta}=\rho_{\gamma\gamma}$, for any 
finite $\Lambda$, and hence population inversion do exist in the 
dressed state basis. For the transitions $|\alpha>\rightarrow |\beta>$,
$|\alpha>\rightarrow |\gamma>$, $|\beta>\rightarrow |\gamma>$
and $|\gamma>\rightarrow |\beta>$ the population difference is negative 
(noninversion) and it remains to be seen whether this transitions 
amplify and thus result in lasing without inversion in the dressed 
state basis. In the strong coupling field limit, $\Omega>>G$ the steady state 
population become
\begin{equation}
\rho_{\beta\beta}=\rho_{\gamma\gamma}=\frac{\gamma_{c}+\Lambda}{2\gamma_{c}+3\Lambda},
\label{16a}
\end{equation}
and
\begin{equation}
	\rho_{\alpha\alpha}=\frac{\Lambda}{2\gamma_{c}+3\Lambda}.
	\label{16b}
\end{equation}

In the following we will get into more detail 
regarding gain without inversion in the dressed state basis.

\paragraph{Evolution of coherences.} Ignoring the ``non secular'' 
couplings between coherences and populations in Eqs. (\ref{12a}) - 
(\ref{12f}) results in equations that are simpler than the original 
ones, 
however, they are still very complicated, particularly the equations for 
$\rho_{\beta\gamma}$ and it's conjugate, which are coupled to all the 
other coherences. Hence one would like to further approximate these 
equations in such a way that the resulting solutions will be fairly 
simple on one hand, and be a reasonable approximation to the exact 
solution on the other. We solved numerically the complete set 
(\ref{12a}) -(\ref{12f}) and found that $\rho_{\beta\gamma}$ and 
hence its conjugate are substantially larger than the other 
coherences, indicating the crucial role these coherences  play. In 
light of the above, we couple each atomic coherence to itself (describing 
the free evolution) and to $\rho_{\beta\gamma}$ and 
$\rho_{\gamma\beta}$ acting as the dominant source terms.

 This gives the following equations:

\begin{mathletters}
\begin{eqnarray}
	\dot\rho_{\alpha\beta} & = & -(\Gamma_{\alpha\beta}-iR)\rho_{\alpha\beta}-
	\tilde{\Gamma'}\rho_{\beta\gamma}+(\tilde{\Gamma}-\tilde{\Gamma'})\rho_{\gamma\beta},
	\label{17a}\\
\dot\rho_{\alpha\gamma} & = & -(\Gamma_{\alpha\gamma}+i 
R)\rho_{\alpha\gamma}+(\tilde{\Gamma}-\tilde{\Gamma'})\rho_{\beta\gamma}-\tilde{\Gamma'
}\rho_{\gamma\beta},
	\label{17b}  \\
	\dot \rho_{\beta\gamma} & = & 
	-(\Gamma_{\beta\gamma}+2iR)\rho_{\beta\gamma}-(\Gamma_{\beta}+\frac{1}{2}
	\Lambda-\frac{1}{2}\Lambda')\rho_{\gamma\beta},
	\label{17c}
\end{eqnarray}
\end{mathletters}

along with the equation for $\rho_{\gamma\beta}=\rho_{\beta\gamma}^{*}$.
Solving the eigenvalue problem of eq.'s (\ref{17a}) - (\ref{17c}) we 
find the transient solutions for the coherences , in the strong 
coupling field limit. These solutions are:

\begin{mathletters}
\begin{eqnarray}
\rho_{\alpha\beta} & = &A\exp[-(\Gamma_{\alpha\beta}-iR)t]+
	C\frac{(\tilde{\Gamma}-\tilde{\Gamma'})(\Gamma_{\beta}+\Lambda/2-\Lambda'/2)
+i4\tilde{\Gamma'}R}{(\Gamma_{\beta}+\Lambda/2-\Lambda'/2)(\Gamma_{\beta\gamma}
-\Gamma_{\alpha\beta}+3iR)}\exp[-(\Gamma_{\beta\gamma}+2iR)t] \nonumber \\
 &+&
D\frac{(\tilde{\Gamma'}-\tilde{\Gamma})+i\tilde{\Gamma'}(
\Gamma_{\beta}+\Lambda/2-\Lambda'/2)/4R}{\Gamma_{\beta\gamma}
-\Gamma_{\alpha\beta}-iR}\exp[-(\Gamma_{\beta\gamma}-2iR)t],
	\label{18a} \\
\rho_{\alpha\gamma}& = &  B\exp[-(\Gamma_{\alpha\beta}+iR)t]+
	C\frac{\tilde{\Gamma'}(\Gamma_{\beta}+\Lambda/2-\Lambda'/2)-i4R(\tilde{\Gamma
	}-\tilde{\Gamma'})}{(\Gamma_{\beta}+\Lambda/2-\Lambda'/2)
	(\Gamma_{\beta\gamma}-\Gamma_{\alpha\beta}+iR)}\exp[-(\Gamma_{\beta\gamma}+2iR)t]
	\nonumber \\
	&+& D\frac{\tilde{\Gamma'}-i(\tilde{\Gamma}-\tilde{\Gamma'})
(\Gamma_{\beta}+\Lambda/2-\Lambda'/2)/4R}{\Gamma_{\beta\gamma}-\Gamma_{\alpha\beta}
-3iR}\exp[-(\Gamma_{\beta\gamma}-2iR)t],
\label{18b} \\
\rho_{\beta\gamma} & = & C\frac{4iR}{\Gamma_{\beta}+\Lambda/2-\Lambda'/2}
\exp[-(\Gamma_{\beta\gamma}+2iR)t]+D\hspace{.2cm}i\hspace{.2cm}
\frac{\Gamma_{\beta}+\Lambda/2-\Lambda'/2}{4R}\times 
\nonumber \\ 
&\times& \exp[-(\Gamma_{\beta\gamma}-2iR)t],
	\label{18c}
\end{eqnarray}
\end{mathletters}

where A, B, C and D are constants ought to be calculated from initial conditions.
Figure 4 shows the exact coherences obtained by solving numerically 
eq.'s (\ref{12a}) -(\ref{12f}) (solid line), and the corresponding 
approximate solutions based on the analytic expressions (\ref{18a}) -(\ref{18c}).
The chosen parameters are the same as in Fig 3.  It can be seen that 
the exact solutions reach a steady state different from zero, while our 
approximate analytic solution have zero steady state values.
This is  due to the absence of the population source terms in 
the approximation.  The 
oscillation frequency is predicted correctly by the 
approximate solutions.  Moreover, the approximate solution for 
$\rho_{\beta\gamma}$ appears to be more accurate than the other 
two, again 
indicating the crucial role played by $\rho_{\beta\gamma}$ and 
$\rho_{\gamma\beta}$.  Note also the negligible contribution of 
$\rho_{\alpha\beta}$ and $\rho_{\alpha\gamma}$ to 
$\rho_{\beta\gamma}$.  The latter coherence is not coupled to 
$\rho_{\alpha\beta}, \rho_{\alpha\gamma}$ yet the approximation remains satisfactory. 
Eq. (\ref{18c}) also shows that $\rho_{\beta\gamma}$ has an almost pure 
sinusoidal form of frequency $2R$. 
To the contrary $\rho_{\alpha\beta}, \rho_{\alpha\beta} $ and $\rho_{\alpha\gamma}$ have a 
composite oscillation, being a superposition of frequencies.
It can be seen from the numerical solution presented in figure 4 
that the coherences $\rho_{\alpha\beta}$, and $\rho_{\alpha\gamma}$ 
possess a definite sign, thus the corresponding transitions between 
dressed states, being either amplified or attenuated. More precisely 
$Im(\rho_{\alpha\beta})$
is positive in all  the range shown, hence the transition 
$|\beta>\rightarrow|\alpha>$ is amplified (with population inversion 
in the dressed state picture at the frequencies $\Omega_{L}+R, 
\Omega_{p}+R$ ).
The transition  $|\alpha>\rightarrow|\gamma>$ is also amplified since 
$Im(\rho_{\gamma\alpha})>0$, however,  without population inversion 
in this case.
\ This situation is very different from that occurring in 
the bare state basis, where the coherences oscillate back and forth 
across zero, thus experiencing periodic amplification and 
absorption see Fig \ref{f4}, and \cite{12}. In contrast to $\rho_{\alpha\beta}$ and 
$\rho_{\alpha\gamma}$, the sign of $Im(\rho_{\gamma\beta})$ is 
alternating thus the transition  $|\beta>\rightarrow|\gamma>$ is 
being amplified and absorbed periodically. Another feature reminiscent 
of figure 4 is the strength of the coherence $\rho_{\beta\gamma}$, 
which is seen to be three orders of magnitude stronger than the other 
two coherences.
The most striking result is that of the transition 
$|\gamma>\rightarrow|\alpha>$. It is absorbing \underline{despite} 
the population inversion (see figure 4 (b) ).
 \paragraph*{}The main deficiency of eq.'s (\ref{18a}) - (\ref{18c}) is the 
zero steady state predicted by them. The reason for this is the 
omission of the source terms (populations) in writing eq.'s 
(\ref{17a}) -(\ref{17c}). We have solved analytically Eqs. (\ref{17a}) 
-(\ref{17c})
with the source terms included. The solution obtained was checked 
against numerical calculations and found to be in excellent agreement. 
Unfortunately, the solution is so complicated, that even reduction 
operations carried out by ``Mathematica'' \cite{25} could not give a 
manageable solution. For the purpose of finding the steady state 
coherences it is sufficient to retain the population terms in (\ref{17a}) -(\ref{17c})
, set to zero the time derivatives, and solve the resulting algebraic 
equations. This yield the following steady state coherences:

\begin{mathletters}
\begin{eqnarray}
	\rho_{\alpha\beta}^{st} & = & (\rho_{\alpha\gamma}^{st})^{*}=
	\frac{(\Gamma_{\alpha\beta}+iR)[\Gamma_{\alpha}\tilde{\Gamma'}+
	(\Gamma_{\alpha}+2\Lambda')(\tilde{\Gamma}+\tilde{\Gamma'})]}
{(\Gamma_{\alpha\beta}^{2}+R^{2})(2\Gamma_{\alpha}+3\Lambda')}\nonumber \\
 & + & \frac{4\Gamma_{\beta}(\Gamma_{\alpha\beta}+iR)(\Gamma_{\alpha}
 +\Lambda')[(2\tilde{\Gamma'}-\tilde{\Gamma})(2\Gamma_{\beta}+\Lambda
 -\Lambda'/2) - 2iR \tilde{\Gamma}]}
 {(\Gamma_{\alpha\beta}^{2}+R^{2})(2\Gamma_{\alpha}+3\Lambda')
 [\Gamma_{\beta\gamma}^{2}+4R^{2}-(\Gamma_{\beta}+\Lambda/2-\Lambda'/2)^{2}]}
	\label{19a},  \\
	& & \nonumber \\
	\rho_{\beta\gamma}^{st} & = & \frac{4\Gamma_{\beta}(\Gamma_{\alpha}+
	\Lambda')}{2\Gamma_{\alpha}+3\Lambda'}\hspace{.3cm}\frac{(2\Gamma_{\beta}+\Lambda
	-\Lambda'/2) -2iR}{(\Gamma_{\beta}+\Lambda/2-\Lambda'/2)^{2}-(
	\Gamma_{\beta\gamma}^{2}+4R^{2})} 
	\label{19b}. 
\end{eqnarray}
\end{mathletters}  
 The steady state values predicted by the last expressions were 
 checked against numerical calculation and found to be in very good 
 agreement. Note that expressions (\ref{19a}) -(\ref
 {19b}) also give 
 the steady state dispersion and not only the gain or absorption 
 coefficients. The expected dominance of 
 $\rho_{\beta\gamma}$ on the other two coherences, can be seen by noting that the 
 term $\tilde{\Gamma}+\tilde{\Gamma'}$, appearing in 
 $\rho_{\alpha\beta}$ and $\rho_{\alpha\gamma}$ (but not in 
 $\rho_{\beta\gamma}$) varies like $ o(G/R)$, thus the coherence 
 $\rho_{\alpha\beta}$ varies like $\rho_{\alpha\beta}\sim G/R^{2}$ 
 while $\rho_{\beta\gamma}$ varies like $\rho_{\beta\gamma}\sim 1/R$. 
 The  second term in $Im(\rho_{\alpha\beta})$ is of the order $G/R^{3}$ 
 and can be neglected. 
 Amplification of the $|\beta>\rightarrow|\alpha>$ transition
 at frequencies $\Omega_{L}+R$, $\Omega_{p}+R$, occurs whenever 
 $Im(\rho_{\alpha\beta}) > 0 $. Taking into account only the first 
 term in (\ref{19a}) we find gain in the following two cases:
 \begin{enumerate}
 	\item  For any incoherent pump rate $\Lambda$ (even zero), 
 	if $\gamma_{b}>\gamma_{c}$. The physical interpretation of this 
 	result is clear: if  level $|b>$ is drained more quickly than $|c>$
 	there is no need for the incoherent pump, and the recycling of the 
 	population is accomplished by the coherent probe field.
 	\item  For $ \frac{\Omega^{2}}{\Omega^{2}+R^{2}} 
 	\gamma_{c}<\gamma_{b}<\gamma_{c}$ provided that the incoherent pump 
 	rate is strong enough such that $\Lambda > 
 	\frac{\Gamma_{\alpha}(\gamma_{c}-\gamma_{b})}
 	{\Gamma_{\alpha}-2\Omega^{2}(\gamma_{c}-\gamma_{b})/R^{2}}$.
 \end{enumerate}
 
This gain is ``regular'' gain, due to population inversion since 
$\rho_{\beta\beta} > \rho_{\alpha\alpha}$. The transition $|\alpha> \rightarrow
|\gamma>$ at frequencies $\Omega_{L}+R$, $\Omega_{p}+R$ will be amplified
 for the same range of parameters because 
the corresponding gain coefficient is proportional to 
$Im(\rho_{\gamma\alpha}) =-Im(\rho_{\alpha\gamma})=Im(\rho_{\alpha\beta})$.
However this transition is inversionless, and it is due to external 
field induced quantum interferences and atomic coherences.
The opposite transition,
$|\gamma>\rightarrow|\alpha>$  is at frequencies $\Omega_{L}-R$, $\Omega_{p}-R$.
It exhibits absorption with population inversion (see figure 4 (b)).
In a sense this is the reverse process of amplification without inversion 
and it is explained as a constructive quantum interference for the 
stimulated absorption process. 
 The  $|\beta> \rightarrow |\gamma>$ transition,  at 
 frequencies $\Omega_{L}+2R$, $\Omega_{p}+2R$ 
will be absorbed, since the term $4R^{2}$ appearing in the 
denominator of (\ref{19b}) far exceeds the other denominator terms 
, resulting in $ Im (\rho_{\gamma\beta}) <0 $, and hence absorption.
The transition $|\gamma>  \rightarrow |\beta>$ at frequencies 
$\Omega_{L}-2R$, $\Omega_{p}-2R$ in turn will be 
amplified for any incoherent pump rate.

In the strong field limit the imaginary parts of the steady state 
coherences can be expressed in terms of the original atomic 
parameters as 

\begin{mathletters}
\begin{eqnarray}
	Im(\rho_{\alpha\beta}^{st}) & = &	Im(\rho_{\gamma\alpha}^{st}) = G 
	\frac{(\gamma_{c}+2\Lambda)(\gamma_{b}-\gamma_{c})+\gamma_{c}\Lambda}{2\sqrt{2}
	\Omega^{2}(2\gamma_{c}+3\Lambda)} ,
	\label{20a}  \\
	 Im(\rho_{\beta\gamma}^{st}) & = & 
	 \frac{\gamma_{b}(\gamma_{c}+\Lambda)}
	 {2\Omega (2\gamma_{c}+3\Lambda)}.
	\label{20b}
\end{eqnarray}€
\end{mathletters} 

As mentioned before, in the strong coupling field limit (week probe) 
the state $|\alpha>$ becomes the highest excited bare state $|c>$, thus 
we are facing a situation of full {\it{noninversion}} in the dressed state 
basis, i.e., $\rho_{cc} < \rho_{\beta\beta}= \rho_{\gamma\gamma}$ (see 
Eq.. (\ref{16a}) - (\ref{16b})). Eq (\ref{20a}) indicates that gain 
can be obtained for the $|c>\rightarrow|\beta>$ and $|c>\rightarrow|\gamma>$ 
transitions (the Autler- Townes transitions \cite{26} ) for the 
following conditions:

\begin{enumerate}
	\item  For any incoherent pump rate $\Lambda$ if 
	$\gamma_{b}>\gamma_{c}$.
	\item  For $ \frac{1}{2}\gamma_{c}<\gamma_{b}<\gamma_{c}$, provided 
	that 
	$\Lambda>\frac{\gamma_{c}(\gamma_{c}-\gamma_{b})}{2\gamma_{b}-\gamma_{c}}$.
\end{enumerate}
This gain is {\it{without inversion}}.
The physics involved in this condition is as follows:\\
It states that the dephasing time of level $|c>$ namely 
$(\frac{1}{2}\gamma_{c})^{-1}$ must be longer than any other decay 
time in the system. The dephasing process must be slow with respect to 
other processes in order to preserve the phase of the dipole 
transition $\rho_{ac}$. This makes possible the quantum coherent 
effect whereby the interference can result in gain without inversion.

\paragraph*{}€To improve the temporal results for the populations we retained only 
the dominant coherences, namely the approximate solutions for 
 $\rho_{\beta\gamma}$ and $
\rho_{\gamma\beta}$ in the population equations (Eq. (\ref{13a})-
(\ref{13c}) serving as 
source terms. These equations are integrated giving the following:
\begin{mathletters}
\begin{eqnarray}
 	\rho_{\alpha\alpha}(t) & = &\rho_{\alpha\alpha}
 	^{st}+c_{1}e^{-(\Gamma_{\alpha}+\frac{3}{2}\Lambda')t}
 	+\rho_{\alpha\alpha}^{par}(t) , \label{21a} \\
 \rho_{\beta\beta}(t) & = &\rho_{\beta\beta}^{st}- \frac{1}{2}
 c_{1}e^{-(\Gamma_{\alpha}+\frac{3}{2}\Lambda')t}+
 c_{2}e^{-(2\Gamma_{\beta}+\Lambda-\frac{1}{2}\Lambda')t}
 -\frac{1}{2}\rho_{\alpha\alpha}^{par}(t),\label{21b} \\   
  \rho_{\gamma\gamma}(t) & = & \rho_{\gamma\gamma}^{st}- \frac{1}{2}
 c_{1}e^{-(\Gamma_{\alpha}+\frac{3}{2}\Lambda')t}-
 c_{2}e^{-(2\Gamma_{\beta}+\Lambda-\frac{1}{2}\Lambda')t}
 -\frac{1}{2}\rho_{\alpha\alpha}^{par}(t).\label{21c}  
 \end{eqnarray}
 \end{mathletters}
 where the particular solution is given by:
 \begin{eqnarray}
 \rho_{\alpha\alpha}^{par}(t) =	-\frac{\rho_{\beta\gamma}(0)\Lambda'\left[
 	(\Gamma_{\alpha}+\frac{3}{2}\Lambda'-\Gamma_{\beta\gamma}
 	)\cos(2Rt)+2R\sin(2Rt)\right]}{(\Gamma_{\alpha}+\frac{3}{2}\Lambda'-\Gamma_{\beta\gamma}
 	)^{2}+4R^{2}}\hspace{.3cm}e^{-\Gamma_{\beta\gamma}t}€
 	\label{€}\nonumber
 \end{eqnarray}
 
 The integration constants are given in terms of initial populations 
 and coherences by:
 \begin{eqnarray*}
  c_{1}&=&\rho_{\alpha\alpha}(0)-\rho_{\alpha\alpha}^{st}-
  \rho_{\alpha\alpha}^{par}(0)\\
   c_{2}&=&\rho_{\beta\beta}(0)-\rho_{\beta\beta}^{st}+
   \frac{1}{2}(\rho_{\alpha\alpha}(0)-\rho_{\alpha\alpha}^{st})
   \end{eqnarray*}
 Figure \ref{f5} shows the difference between  the exact population
 $\rho_{\alpha\alpha}$ and the approximate solution 
 (\ref{21a})-(\ref{21c}). It can be seen that the approximate solution is 
 very accurate.

 To compare with the steady state situation in the bare 
state basis, we need to transform back to the bare state basis ,via 
the the matrix product $\rho^{B}=T^{-1}\rho^{Dr}T$, where $\rho^{Dr}$ 
is the density matrix in the dressed state basis, formed by the 
steady state populations and coherences of Eq..'s (\ref{15a}) 
-(\ref{15b}), and (\ref{19a}) -(\ref{19b}). Further utilization of the 
strong coupling field limit gives  bare state populations, in the 
steady state regime.

\begin{mathletters}
\begin{eqnarray}
	\rho_{aa} & = & 
	\rho_{bb}=\frac{\gamma_{c}+\Lambda}{\gamma_{c}+3\Lambda},
	\label{22a}  \\
\text{ and }\nonumber \\
	\rho_{cc} & = & \frac{\Lambda}{\gamma_{c}+3\Lambda}.
	\label{22b}
\end{eqnarray}€
\end{mathletters}

The bare state coherences are:

\begin{mathletters}
\begin{eqnarray}
	\rho_{ab} & = & -i 
	\frac{\gamma_{b}(\gamma_{c}+\Lambda)}{2\Omega(2\Gamma_{c}+3\Lambda)},
	\label{23a}  \\
	 \rho_{ac} & = & i 
	 \hspace{.1cm}G\frac{\Lambda(\gamma_{b}-\gamma_{c})-\gamma_{c}^{2}}
	 {2\Omega^{2}(2\gamma_{c}+3\Lambda)},
	\label{23b}\\
	\text{and}\nonumber \\
	\rho_{bc} & = & \frac{G\gamma_{c}}{\Omega (2\gamma_{c}+3\Lambda)}.
	\label{23c} 
\end{eqnarray} 
\end{mathletters}
In obtaining the expressions for populations and coherences we find 
that our general form reduce to previously obtained results \cite{12}. 

We see that in the bare state basis one always has $Im(\rho_{ab})<0$, 
thus the coupling laser is always attenuated. The probe transition 
exhibits inversionless gain for $\gamma_{b}>\gamma_{c}$ for pump 
rates satisfying $\Lambda>\frac{\gamma_{c}^{2}}{\gamma_{b}-\gamma_{c}}$.
From the analysis presented above we conclude that for a week probe 
, true lasing without population inversion can be realized, both in 
the bare state and the dressed state basis. 

\section{Summary} 
Absorption in the presence of inversion and amplification without 
inversion in a three level V - type system are found in the dressed state 
picture. Both of these effects are the manifestation of the quantum 
interference that occurs in multilevel systems. Moreover, the above two 
processes constitute a manifestation of a complementarity principle.
\paragraph*{} We have presented an analysis of light amplification 
without population inversion in this system within 
the framework of the dressed state basis. The equations of motion for the 
elements of the density matrix are derived from the 
master equation. In the dressed state picture new relaxation terms 
are defined that are related dierctly to the coherently prepared states 
and the quantum interfernce effects.  interferference terms are 
identified. They are shown to be the source for amplification without 
inversion and for absorption despite the inversion.
Consequently, approximate analytical time 
dependent solutions for dressed state populations and coherences were obtained. 
Comparison of these approximate solutions with the numerically calculated 
quantities shows excellent agreement. Both of these solutions exhibit the 
familiar Rabi oscillations. Steady state density matrix elements 
were also calculated, from which we have concluded that for a weak 
probe field, true lasing without inversion exist for the appropriate
incoherent pump rate, i.e., lasing without inversion in any state 
basis. Steady state quantities were transformed back to the bare state basis, 
and were found to be in perfect agreement with results in the literature.  
Conditions for inversionless gain where obtained and show physically 
sound basis. 
Finally, the new feature of absorption despite population inversion found in 
this calculation emphasizes the importance of quantum 
interference.  Quantum interference is thus shown to imprint its effect 
on the processes in the dressed state picture.

\newpage
\newpage
\begin{center}
\begin{figure}[p]
	\centering
	\includegraphics{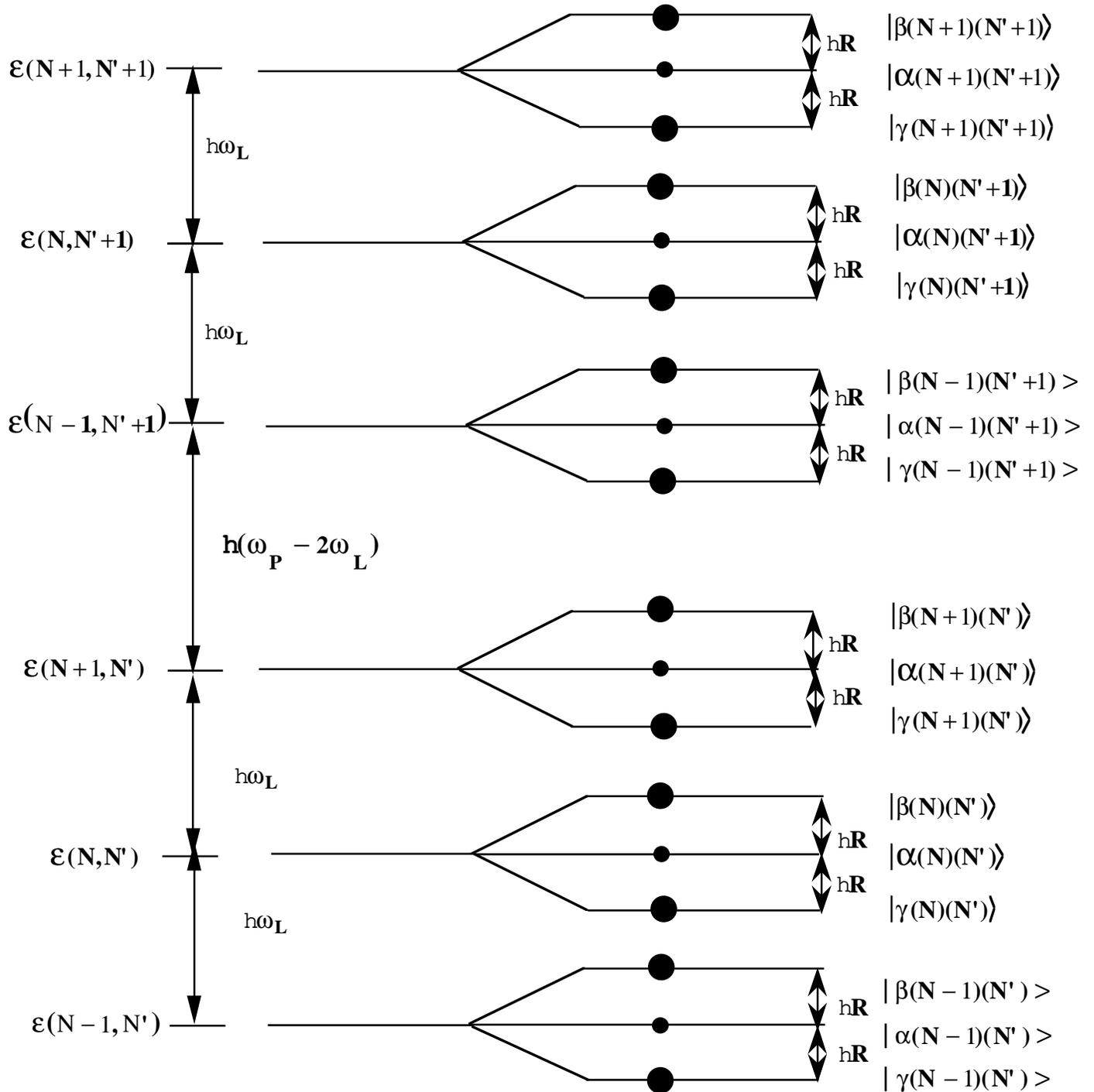}
\caption{Manifolds $\epsilon(N,N'), \epsilon(N+1,N'),\epsilon(N,N'+1)$, 
and $\epsilon(N+1,N'+1)$ etc. of uncoupled states of the atom + lasers 
photons (left hand part). The dressed level (perturbed levels) are shown 
at the right hand side. The circles represent steady state populations.  }
	\label{f2}
\end{figure}
\end{center}
\newpage
\begin{center}
\begin{figure}[b]
	\centering
	\includegraphics[50,590][530,740]{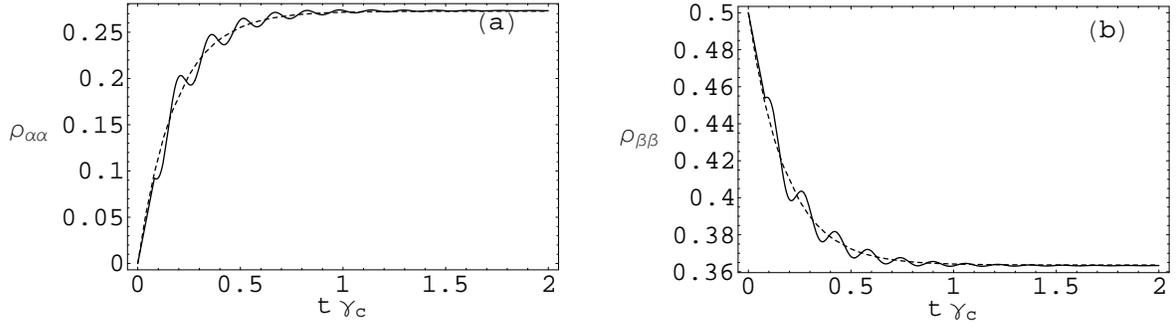}
	\vspace{1cm}
\caption{ Normalized time evolution numerical simulation, of dressed state  
population $\rho_{\alpha\alpha}$ and $\rho_{\beta\beta}$ obtained by solving Eq.s (\ref{12a}) - (\ref{12f}) 
(solid line), and the approximate solution based on Eq.s 
(\ref{14a}) -(\ref{14c})(dashed line)
.The chosen parameters are: $\Omega=20\gamma_{c}$, $\gamma_{b}=2\gamma_{c}$, $G=0.1\gamma_{c}$, 
 $\Lambda=3\gamma_{c}$}
	\label{f3}
\end{figure}
\end{center}
\newpage
\begin{center}
\begin{figure}[p]
	\centering
	\includegraphics[73,284][534,706]{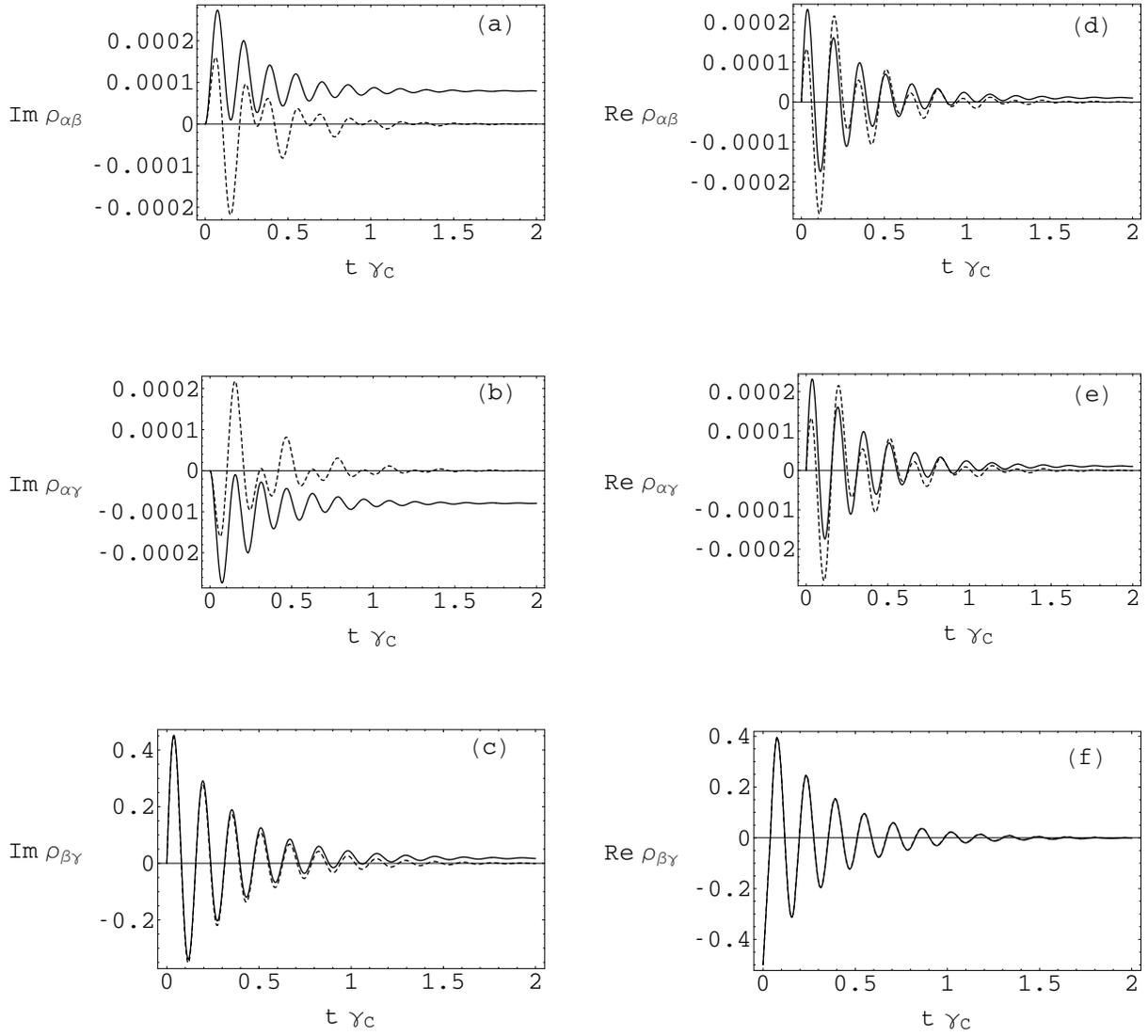}
	\vspace{1cm}
\caption{Normalized time evolution numerical simulation, of dressed state  
coherences obtained by solving Eq.'s (\ref{12a}) - (\ref{12f}) 
(solid line), and the approximate solution based on Eqs. 
(\ref{18a}) -(\ref{18c}) (dashed line). The chosen parameters are the same as in Fig 
\ref{f3}}. Note the absorption despite population inversion seen in (b).
	\label{f4}
\end{figure}
\end{center}
\begin{center}
\begin{figure}[p]
	\centering
	\includegraphics[37,284][480,742]{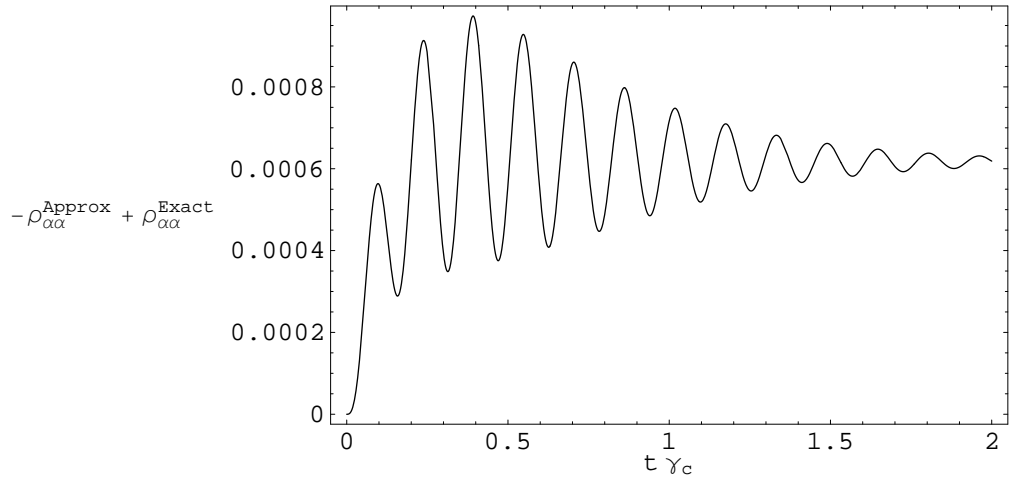}
	\vspace{1cm}
\caption{ Difference between Exact population $\rho_{\alpha\alpha}$ 
obtained by solving Eqs. (\ref{12a}) - (\ref{12f}) 
and the approximate solutions based on  solutions 
(\ref{21a}-\ref{21c}).}
	\label{f5}
\end{figure}
\end{center}
\end{document}